\documentclass[%
 reprint,
 amsmath,amssymb,
 aps,
]{revtex4-2}

\usepackage{graphicx}
\usepackage{dcolumn}
\usepackage{bm}
\usepackage{caption}
\usepackage{subcaption}
\usepackage{graphicx}
\usepackage{amsmath}
\usepackage{booktabs}

\begin{document}

\title{Event generation and consistency tests with sliced Wasserstein distance in high-energy physics}

\author{Chu-Cheng Pan}
\altaffiliation{These authors contributed equally to this work.}
\author{Xiang Dong}
\altaffiliation{These authors contributed equally to this work.}
\author{Yu-Chang Sun}
\author{Ao-Yan Cheng}
\author{Ao-Bo Wang}
\author{Yu-Xuan Hu}
\author{Hao Cai}
\email{hcai@whu.edu.cn}
\author{Wei Wang}
\email{wangwei2017@whu.edu.cn}
\affiliation{%
School of Physics and Technology, Wuhan University, Wuhan, Hubei, China
}%

\begin{abstract}
In the field of modern high-energy physics research, there is a growing emphasis on utilizing deep learning techniques to optimize event simulation, thereby expanding the statistical sample size for more accurate physical analysis. Traditional simulation methods often encounter challenges when dealing with complex physical processes and high-dimensional data distributions, resulting in slow performance. To overcome these limitations, we propose a solution based on deep learning with the sliced Wasserstein distance as the loss function. By employing an advanced transformer learning architecture, we initiate the learning process from a Monte Carlo sample, and generate high-dimensional data. Through the integration of the SWD with the permutation test, we introduce a novel, statistically rigorous and more sensitive metric for assessing the distribution differences, which significantly outperforms other metrics in detecting subtle distributional shifts, further validating its effectiveness for precise evaluation in high-energy physics generative models and high-dimensional consistency test. The generated data samples maintain all the original distribution features from a limited number of training samples, as evidenced by their successful passage of all common consistency tests using a test‑sample size of the same order of statistical magnitude. This development opens up new possibilities for improving event simulation and high-dimensional consistency test in high-energy physics research.
\end{abstract}


\maketitle


\section{Introduction}
In modern high-energy physics research, experimental data are increasingly complex and high-dimensional. For instance, a single particle collision event can generate numerous primary particles, each characterized by parameters such as momentum, energy, charge, and mass \cite{besIII_detector, lhcb_detector}. Monte Carlo simulation is commonly employed to generate particle collisions. But as the complexity of high-energy particle physics detectors grows, so does the intricacy of the detector simulation programs, resulting in higher costs. This issue will be further exacerbated in future high-energy physics experiments like HL-LHC/CEPC/super-tau-charm \cite{future_accelerator, challenges_hl_lhc}. The demand for simulation data will continue to rise as high-energy physics exploration deepens. Additionally, complex data necessitate sophisticated analysis tools, which often require massive storage to obtain a sufficient number of independent simulation data samples. This places significant demands on computing power and storage systems \cite{hard_simulation}. Thus, finding methods to generate high-precision simulation data to reduce computing and storage requirements has become a crucial research objective.

Machine learning has demonstrated immense potential in event generation, with event generators frequently trained using Generative Adversarial Networks (GANs) and Variational Autoencoders (VAEs). The objective is to learn the features of finite statistical data and rapidly generate large amounts of simulated data with consistent features. However, several challenging problems need to be addressed \cite{mleg_survey}. In GANs, the generator aims to simulate the target data, while the discriminator tries to distinguish between the generated data and the target data. Consequently, the training process can be unstable, and issues like model collapse may arise \cite{Di_Sipio_2019}. On the other hand, VAEs map input data to a probability distribution in latent space through an encoder and then reconstruct the original data space using latent variables and a decoder. VAEs may overlook crucial details in the data, necessitating a balance between the reconstruction error and the prior distribution error of the latent space \cite{bvae}. 

To resolve this challenge, we propose the utilization of a novel loss function that captures the overall consistency between the target data and the generated data. The Sliced Wasserstein distance (SWD) presents a potential solution as an effective loss function\cite{swd_theory}. SWD quantifies the disparity between two probability distributions by slicing the probability distribution in multiple directions, calculating the one-dimensional Wasserstein distance in each direction, and averaging the distances across all directions. With its continuous differentiability and high computational efficiency, SWD offers a promising approach to address the loss function problem and optimize model parameters to ensure that the generated data distribution closely approximates the real data distribution \cite{computational_ot}.

Beyond using SWD as a training loss function, there is also growing interest in robustly evaluating generative models using high-dimensional metrics more refined than simple histogram comparisons. Metrics such as the Kernel Physics Distance (KPD) and Fréchet Physics Distance (FPD) \cite{Kansal_2023} have been explored to systematically measure discrepancies between generated and real data. Building on these developments, we combine SWD with a permutation-based statistical test, thereby constructing a rigorous measure of distributional difference. This integrated approach not only offers a straightforward and interpretable way to verify the fidelity of high-dimensional data, but also provides a robust and more sensitive framework for detecting subtle discrepancies that might remain hidden when relying on histograms or other metrics.

\section{Methods}
\subsection{SWD Theory}

Sliced Wasserstein distance (SWD) is a variant of the Wasserstein distance, which transforms the high-dimensional Optimal Transport problem into a one-dimensional problem for ease of computation. Specifically, for each randomly chosen direction, SWD projects the distribution onto this direction and then calculates the Wasserstein distance of these one-dimensional distributions. This process is repeated multiple times and the results are averaged to obtain the final distance estimate. First, let us define the one-dimensional Wasserstein distance. For the one-dimensional case, the Wasserstein distance between two distributions $P$ and $Q$ can be defined as the L1 distance between their cumulative distribution functions (CDF) \cite{wd_two_sample_test}, i.e.,

\begin{equation}
   WD(P,Q) = \int |F_{P(x)} - F_{Q(x)}|dx 
\end{equation}

where $F_P$ and $F_Q$ are the cumulative distribution functions of the distributions $P$ and $Q$. For the high-dimensional case, the Sliced Wasserstein Distance is defined through the integral of the one-dimensional Wasserstein distance:

\begin{equation}
    SWD(P,Q) = \int WD(P_{\theta}, Q_{\theta}) d \theta
\end{equation}

where $P_\theta$ and $Q_\theta$ are the one-dimensional distributions of the P and Q distributions projected in the direction $\theta$. $\theta$ is a direction randomly drawn from a uniform distribution. $W(P_{\theta}, Q_{\theta})$ is the Wasserstein distance between these two one-dimensional distributions. Note that the integral here is over all possible directions $\theta$, i.e., in practice, we usually approximate this integral by sampling a large number of random directions and taking the average. It should be noted that the formula here is for the case of continuous probability distributions. For the discrete case, we assume that we have two sets of points $X={x_1, ..., x_n}$ and $Y={y_1, ..., y_n}$, each set is in $R^d$. We can define the one-dimensional discrete Wasserstein distance as:

\begin{equation}
    WD(X_{\theta}, Y_{\theta}) = \frac 1 n \sum_{i=0}^n x_{i\theta} - y_{\pi(i) \theta}
\end{equation}

where $X_{\theta}$ and $Y_{\theta}$ are the projections of X and Y in the direction $\theta$, and $\pi$ is a permutation that minimizes the above sum. In the one-dimensional case, such $\pi$ can be found by sorting the points after projection. This is because in one dimension, the optimal transport scheme is to pair each point $x_{i\theta}$ with its immediate neighbor $y_{\pi(i)\theta}$. Then, the Sliced Wasserstein Distance for the discrete case is defined as

\begin{equation}
    SWD(X, Y) = E[WD(X_\theta, Y_\theta)] 
\end{equation}

The expectation is calculated by sampling along multiple random directions $\theta$. We choose the projection dimension in the unit circle, which can make the projection more uniform, and then calculate and average the one-dimensional Wasserstein distances in these directions \cite{radon_transfrom, swd_theory}. Some revised theories regarding the Sliced Wasserstein Distance have been proposed to better accommodate complex high-dimensional data. These modifications have potential applications in event generation in future work \cite{gswd, fast_gswd}. 

To correct for detector effects and the spatial defects in high-dimensional data caused by event selection, we will make additional corrections to the dimensions of specific physical quantities, so our total loss is

\begin{equation}
    Loss = SWD(X, Y) + \gamma \cdot E[WD(X_{\theta_{specific}}, Y_{\theta_{specific}})]
\end{equation}

\subsection{Network Architecture and Model Training}
\begin{figure*}
    \centering
    \includegraphics[width=\textwidth]{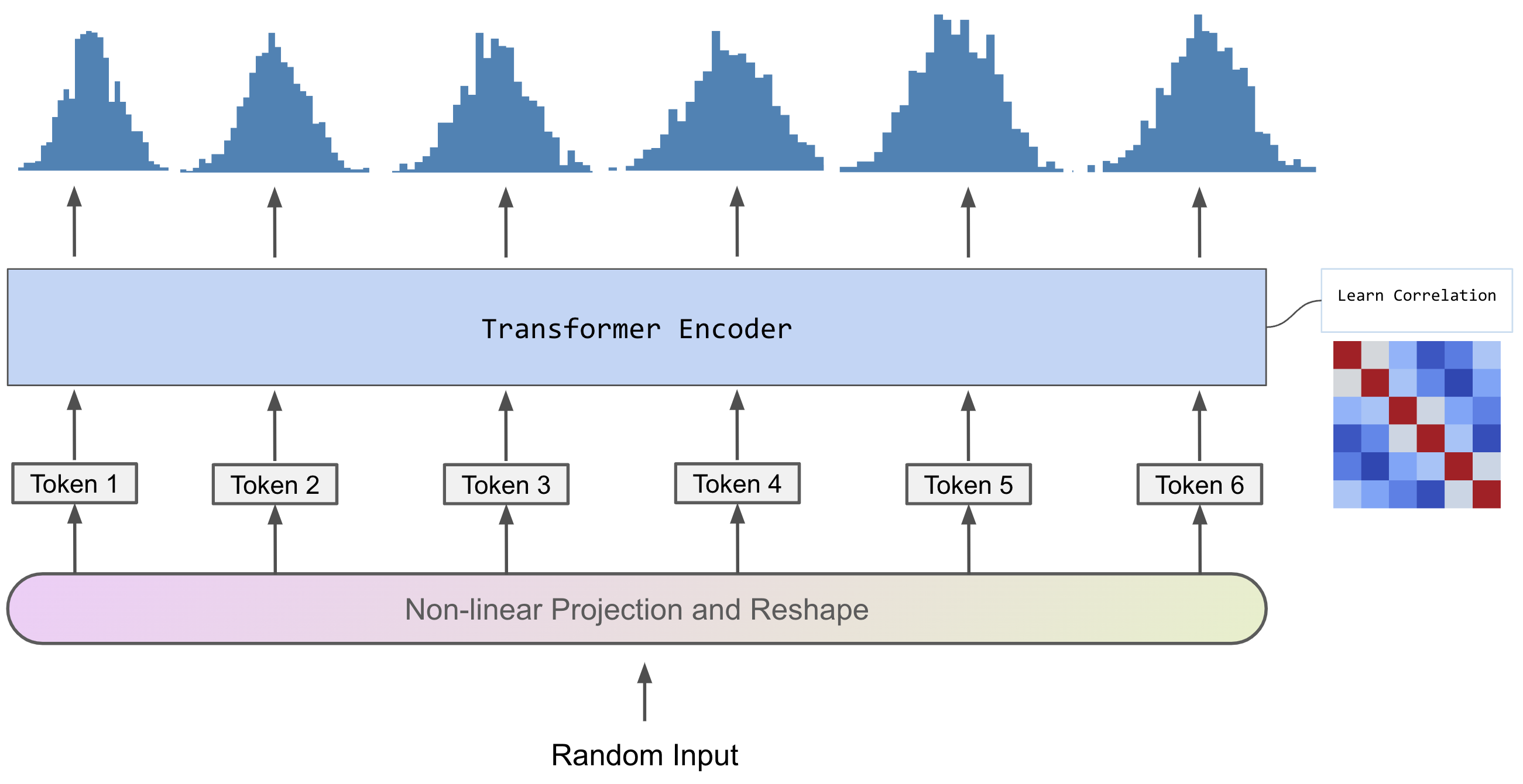}
    \caption{\textbf{Model Overview} We input the random samples from a uniform distribution, non-linearly expand it, split it into desired tokens, and feed it into a standard Transformer encoder to learn the correlation of the multi-variable distribution. Finally, we generate reasonable high-dimensional data.}
    \label{fig:model}
\end{figure*}

To generate complex high-dimensional multivariate distributions, we employ the Transformer model based on the self-attention mechanism \cite{transformer}. The self-attention mechanism allows the model to compute weighted relationships between tokens in an input sequence, effectively capturing intricate interdependencies within high-dimensional data. Our approach begins by generating a latent vector sampled uniformly from -1 to 1, serving as a low-dimensional representation of the distribution. This latent vector is projected through a linear transformation onto a higher-dimensional space corresponding to the number of degrees of freedom in our target distribution, i.e. the number of independent variables or features we aim to model, and scaled by a factor of 1024. The resulting high-dimensional vector is reshaped into a sequence of tokens, each representing an independent unit distribution and containing 1024 elements, which encode information about 1024 points along a particular dimension. By conceptualizing the complex multivariate distribution as a sequence of these unit distribution tokens, we leverage the Transformer’s capability to process sequential data and capture long-range dependencies through self-attention. The Transformer model learns the complex relationships among the different dimensions of the distribution, enabling it to generate new samples that align with the underlying data structure and dependencies. A diagram illustrating the transformations and data flow through the model is provided in Fig. \ref{fig:model}.

We used the AdamW optimizer \cite{loshchilov2019decoupledweightdecayregularization}, with a learning rate set to $5\times10^{-5}$, epsilon set to $10^{-8}$, and weight decay set to 0.1. For each iteration, we randomly extract events for matching and loss calculation, targeting a batch size of 256 by default, and each batch has 1024 events. At the same time, for events where certain physical quantities exceed the prior range, such as detector angles and event selection, we will directly eliminate them during training to exclude events that are in an unreasonable range. The target data will be reduced synchronously, generally slightly fewer than the extracted events.

\section{Results}
We generate phase-space Monte Carlo events and simulate the detector effect using the Geant4-based simulation software developed for the BESIII detector \cite{agostinelli2003geant4, ablikim2010design}, where the $\psi(2S)$ meson is produced through electron-positron collisions and subsequently decays into a $\phi$ meson and a pair of pions ($\pi^+$ and $\pi^-$). The $\phi$ meson further decays into a pair of kaons ($K^+$ and $K^-$) \cite{BaBar:2011btv}. The precision of our generation can be assessed by considering the narrow mass peak of the $\phi$ meson as one of the evaluation metrics. For $K^{+} \pi^{-}$ and $K^{-} \pi^{+}$, we have $M_{K \pi}<0.85, M_{K \pi}>0.95$ to remove possible $K^*$ events. The application of these specific event selection criteria and detector effect aids in evaluating our generation when addressing potential discontinuities in the simulation data. Our simulation data, transformed into the center-of-mass system of $\psi(2S)$, encompass eight degrees of freedom, including the masses $M_{\phi}$ and $M_{\pi^+ \pi^-}$, and the angular distributions of the decay products both in the rest frames of $\phi$ and $\pi^+ \pi^-$. Specifically, we selected the angular variables $\theta_{\phi}$, $\phi_{\phi}$, and the $\theta$ and $\phi$ angles for $K^+$ and $K^-$ in the rest frame of $\phi$, similarly for $\pi^+ \pi^-$, to capture the full dynamics of the decay process.

\begin{figure*}
    \centering
    \begin{subfigure}{.32\textwidth}
        \centering
        \includegraphics[width=\textwidth]{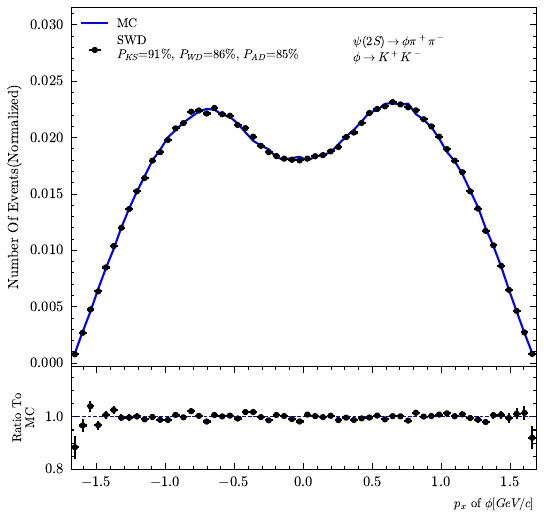}
    \end{subfigure}
    \begin{subfigure}{.32\textwidth}
        \centering
        \includegraphics[width=\textwidth]{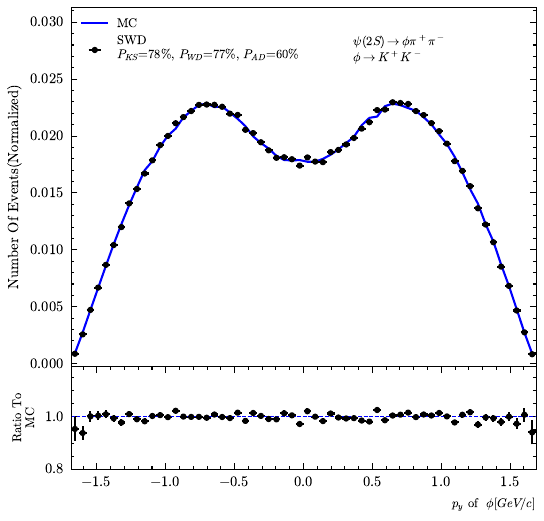}
    \end{subfigure}
    \begin{subfigure}{.32\textwidth}
        \centering
        \includegraphics[width=\textwidth]{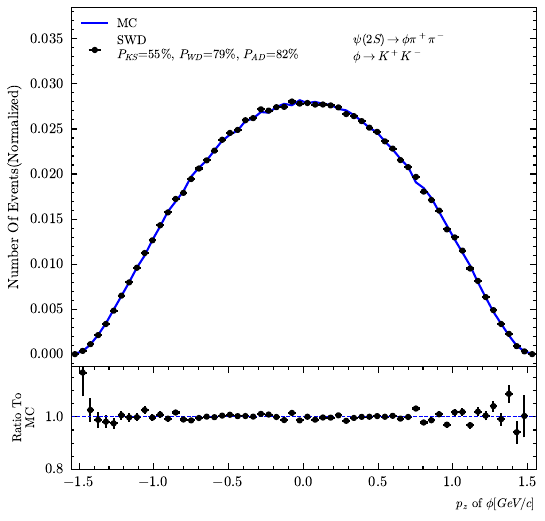}
    \end{subfigure}
    \begin{subfigure}{.32\textwidth}
    \centering
    \includegraphics[width=\textwidth]{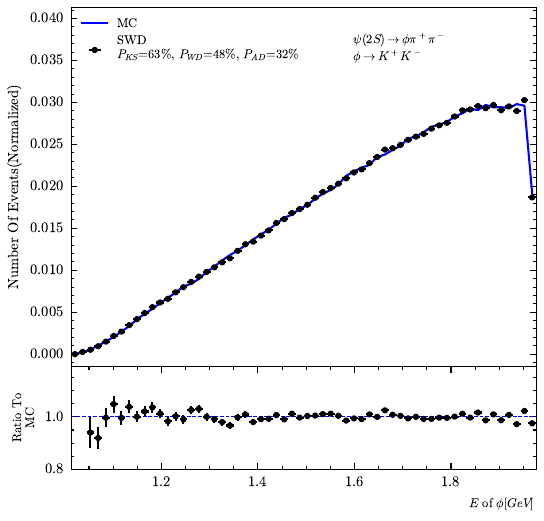}
    \end{subfigure}
    \begin{subfigure}{.32\textwidth}
        \centering
        \includegraphics[width=\textwidth]{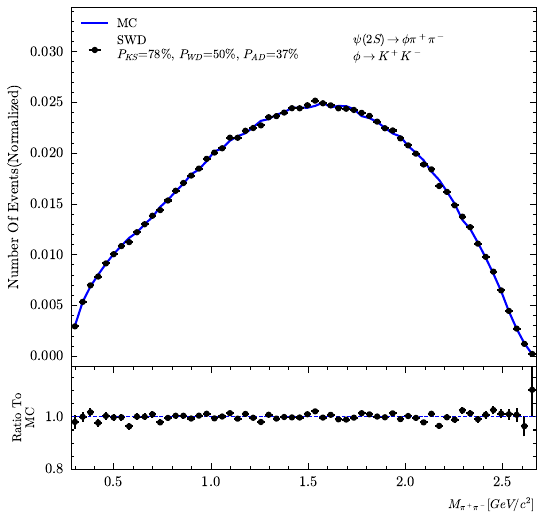}
    \end{subfigure}
    \begin{subfigure}{.32\textwidth}
        \centering
        \includegraphics[width=\textwidth]{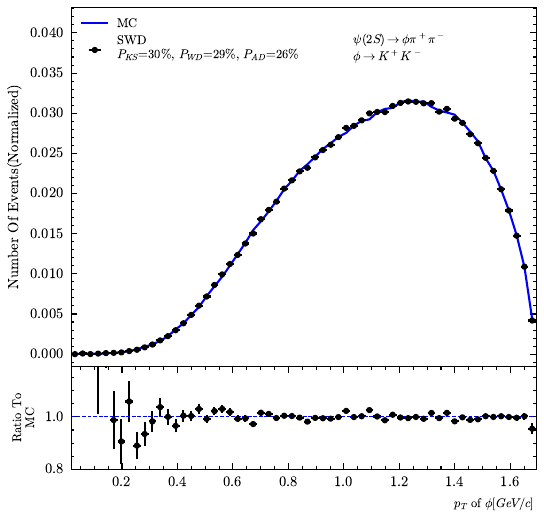}
    \end{subfigure}
    \caption{\textbf{1D Histograms of the kinematic distributions of the $\phi$ meson for one million events} The blue line represents the distribution from Monte Carlo (MC) simulation events, while the black data points are from SWD-generated events. To measure the consistency of these one-dimensional distributions, we employ various statistical tests including the Kolmogorov-Smirnov (KS) test, Wasserstein distance (WD) test, and Anderson-Darling (AD) test.}
    \label{fig:specific projection distributions of phi}
\end{figure*}

\begin{figure*}
    \centering

    \begin{subfigure}{.32\textwidth}
        \centering
        \includegraphics[width=\textwidth]{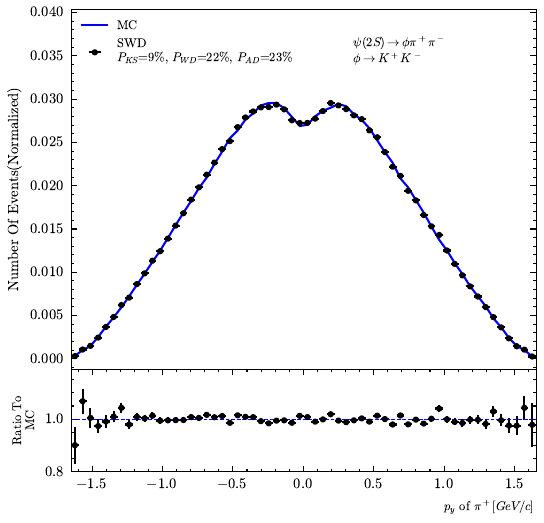}
    \end{subfigure}
    \begin{subfigure}{.32\textwidth}
        \centering
        \includegraphics[width=\textwidth]{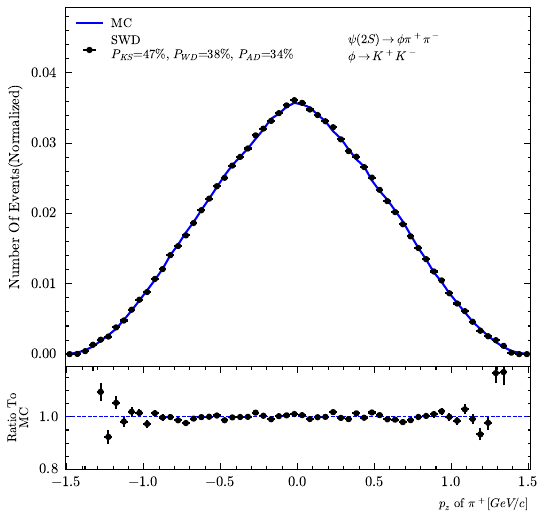}
    \end{subfigure}
    \begin{subfigure}{.32\textwidth}
        \centering
        \includegraphics[width=\textwidth]{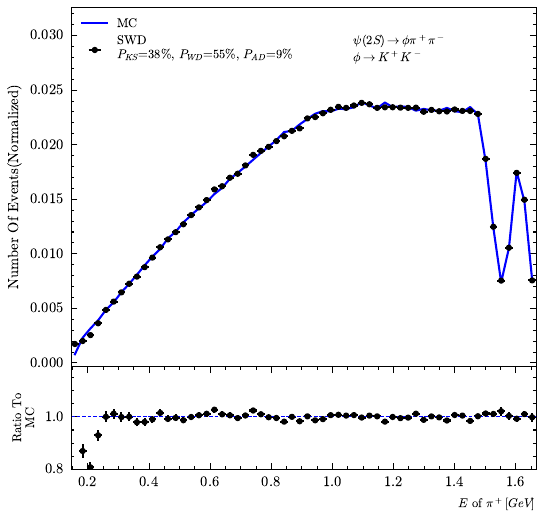}
    \end{subfigure}
    \begin{subfigure}{.32\textwidth}
        \centering
        \includegraphics[width=\textwidth]{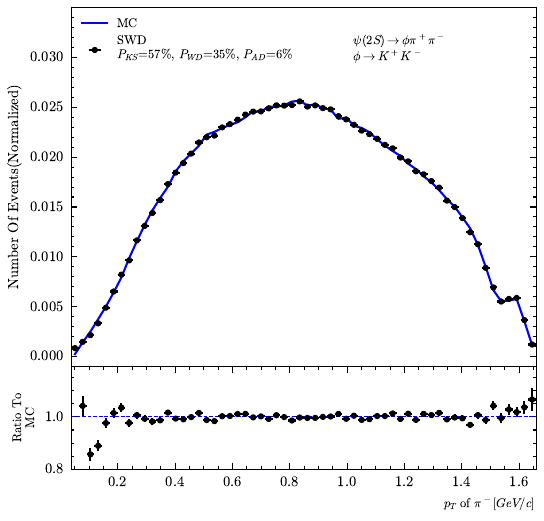}
    \end{subfigure}
    \begin{subfigure}{.32\textwidth}
        \centering
        \includegraphics[width=\textwidth]{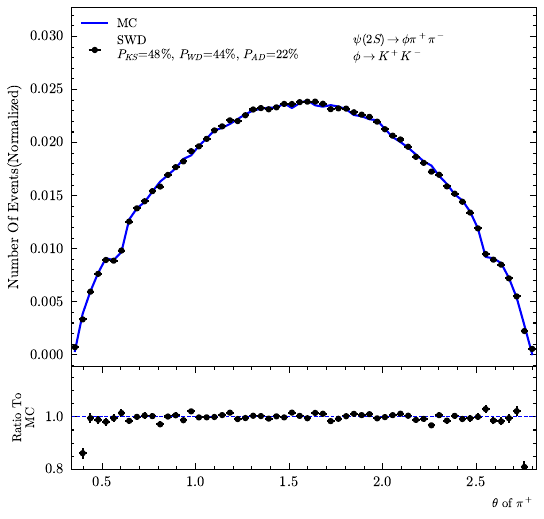}
    \end{subfigure}
    \begin{subfigure}{.32\textwidth}
        \centering
        \includegraphics[width=\textwidth]{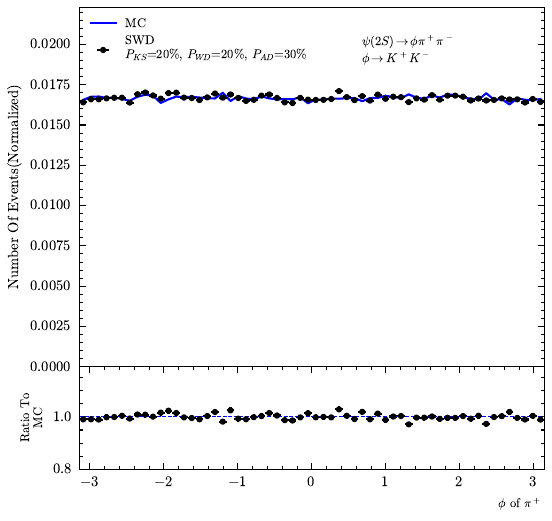}
    \end{subfigure}
    \begin{subfigure}{.32\textwidth}
        \centering
        \includegraphics[width=\textwidth]{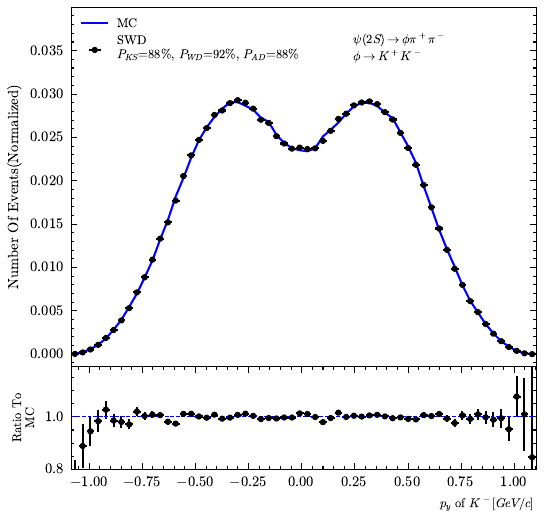}
    \end{subfigure}
    \begin{subfigure}{.32\textwidth}
        \centering
        \includegraphics[width=\textwidth]{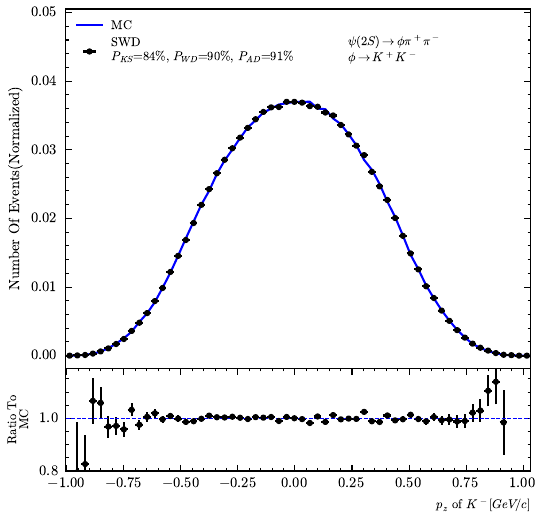}
    \end{subfigure}
    \begin{subfigure}{.32\textwidth}
        \centering
        \includegraphics[width=\textwidth]{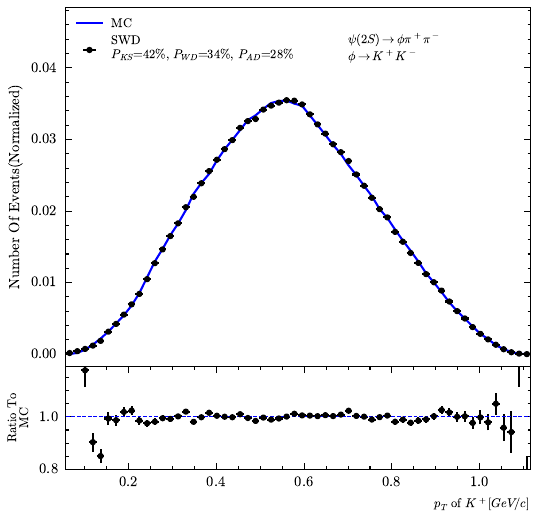}
    \end{subfigure}
    \caption{\textbf{1D Histograms of the kinematic distributions of the final-state particles for one million events} The blue line represents the distribution from Monte Carlo (MC) simulation events, while the black data points are from SWD-generated events. To measure the consistency of these one-dimensional distributions, we employ various statistical tests including the Kolmogorov-Smirnov (KS) test, Wasserstein distance (WD) test, and Anderson-Darling (AD) test. Notably, the oscillatory effect observed in the $E\left(\pi^{+}\right)$distribution near 1.5 GeV is attributed to the veto of $K^*$ events.}
    \label{fig:specific projection distributions}
\end{figure*}

To verify the consistency between the distributions of the simulated data generated by SWD using 2 million training events and the Monte Carlo test set data, we used the Kolmogorov-Smirnov (KS) test, the Wasserstein Distance (WD) and Anderson–Darling (AD) test to conduct a one-dimensional same-distribution test both using 1 million test events in Fig.\ref{fig:specific projection distributions of phi} and Fig.\ref{fig:specific projection distributions}. On specific physical quantity dimensions, the confidence results confirmed that our SWD-generated data and the test set data have high consistency in the distribution on these dimensions.

\begin{table}[h!]
\caption{Percentage of p-values obtained for various statistical metrics when comparing data generated by the SWD method to the test set data, using a sample size of 1,000,000 events.}
\label{table: evaluation metric}
\begin{tabular}{lc}
\hline
\textbf{Metric} & \textbf{P-value} \\
\hline
FPD & 99.9\% \\
KPD & 82.0\% \\
SWD & 24.7\% \\
\hline
\end{tabular}
\end{table}

To accurately assess the differences between high-dimensional distributions \cite{ahmad2024comprehensive, das2024understand}, we combined the SWD and permutation test methods. Firstly, for two high-dimensional distributions, the original distance between them is calculated using the SWD. To create a reference distribution, we merged the datasets $X$ and $Y$ and randomly reorganized them multiple times to generate a series of new data pairs ($X_i'$, $Y_i'$), noting that this reorganization is not based on the original grouping labels. Subsequently, for each pair of newly matched datasets $X_i'$ and $Y_i'$, we calculate the SWD between them to obtain a series of permutation distances ($D_{\text{perm}_1}$, $D_{\text{perm}_2}$, ..., $D_{\text{perm}_n}$). In the evaluation process, we compare the position of the original distance ($D_{\text{original}}$) in these permutation distances. For example, if $D_{\text{original}}$ is greater than 95\% of the $D_{\text{perm}_i}$ distances, a p-value of 0.05 can be obtained. If the p-value is less than 5\%, we reject the null hypothesis, indicating that the samples are likely not from the same distribution. By combining the SWD with the permutation test, we provide a statistically rigorous method for assessing the differences between high-dimensional distributions, which is amenable to GPU acceleration and parallelization. We applied our method and other metrics like the Fréchet and kernel physics distances (FPD and KPD) which are designed as analogs to the commonly used Fréchet Inception Distance (FID) and Maximum Mean Discrepancy (MMD) in machine learning, but applied directly to physics feature spaces \cite{Kansal_2023}, to test our SWD-generated data against the test set data in Table. \ref{table: evaluation metric} and it suggests that our generated data can be considered as originating from the same distribution as the Monte Carlo simulation data.

Having found no significant differences between our generated sample and Geant4 simulations, we proceed to compare the computational performance of our approach with the standard Geant4 pipeline. Training our model requires 12 hours and 26 minutes on a single GPU (GeForce RTX 3090). Once trained, the model can generate 50000 events in just one second. In stark contrast, the standard Geant4 simulation simulates approximately 0.2 events per second. Our method is several orders of magnitude faster than traditional simulation methods. Considering that experimental analyses often require the simulation of billions of simulation events, due to an escalating volume of experimental data, low statistical efficiencies, and the necessity to generate multiple Monte Carlo datasets for the pull distribution, the computational storage savings offered by our method are substantial.

\section{Discussion}
\subsection{Support Size}

\begin{figure}
    \includegraphics[width=0.45\textwidth]{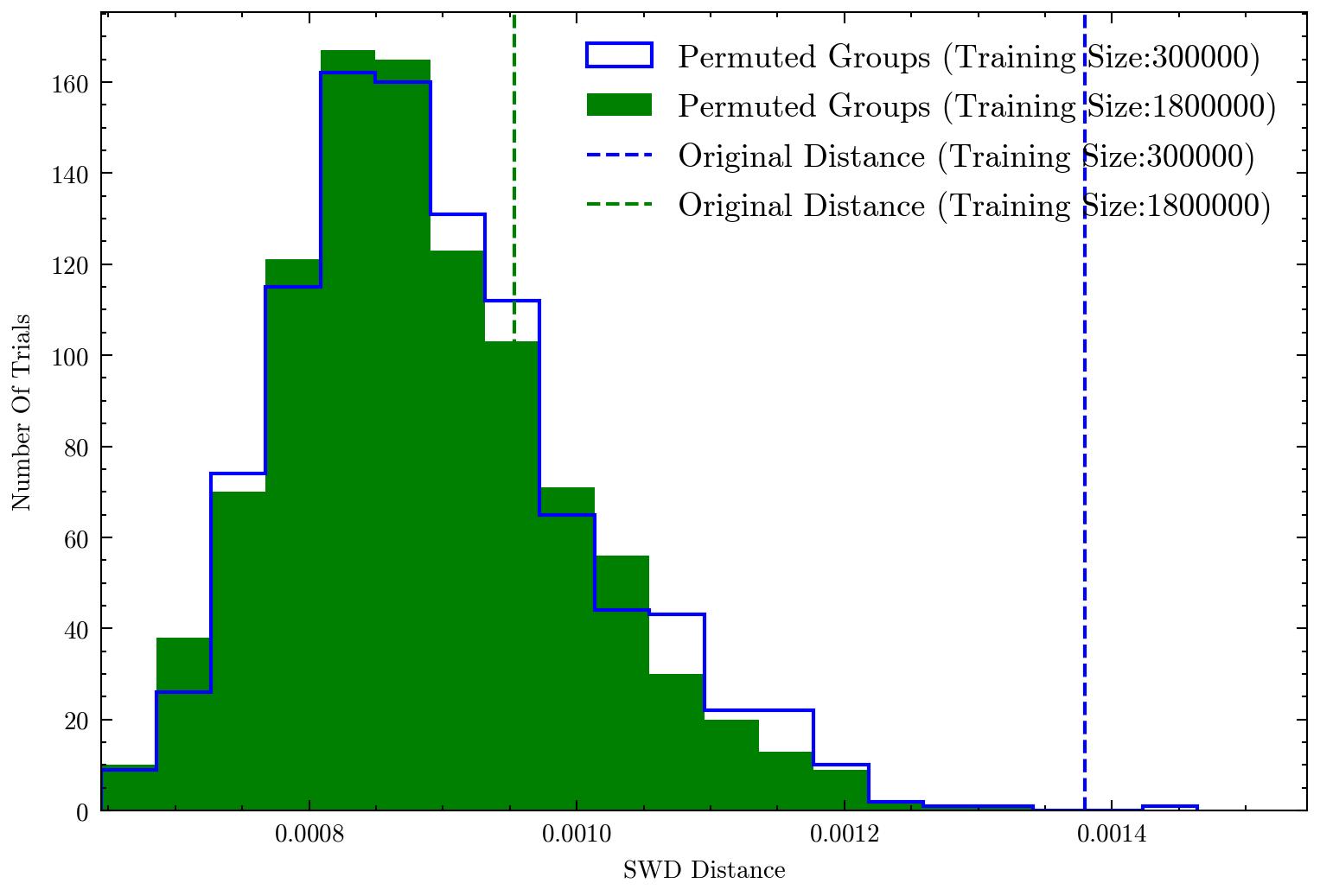}
    \caption{\textbf{Permutation Test} We illustrates the SWD distance distributions obtained via the permutation test for two distinct training sample sizes (300,000 vs. 1.8 million events). Each bar corresponds to the frequency of SWD distances generated through random permutations of the data. The vertical dashed lines mark the observed SWD distances from the unpermuted datasets}
    \label{fig:permutation_test_example}
\end{figure}

In high-energy physics generative modeling, performance often depends on the number of training samples. Consequently, understanding the relationship between training sample size and support test sample size is important. Figure \ref{fig:permutation_test_example} illustrates the permutation test distribution both using 1 million test events. The null hypothesis posits that the two samples originate from identical distributions. The p-value can be approximated by the ratio of distances within the permutation groups that exceed the original distance. The null hypothesis cannot be rejected due to its p-value of 26.6 \% when the model is trained with 1.8 million events. However, the null hypothesis can be rejected given that the p-value is near 0 \% when the training size is 300,000. 

Suppose a specific physics analysis requires generating 1 million events, a natural question arises as to how large the training dataset must be to ensure statistical robustness. Table 
 \ref{table of train sample} presents the influence of the training sample size on the performance of the algorithm using 1 million test events. From the results, it is evident that increasing the training sample size enhances the model's ability to learn, as reflected by the reduction in the SWD test distance and an increase in the p-value for larger sample sizes. However, beyond a certain point like 1 million training events, further increases in the training sample size yield diminishing returns in terms of model learning improvement. Specifically, we find that when the training sample size exceeds approximately 750,000 events, the SWD permutation test p-value exceeds 5\%, suggesting that this sample size is sufficient for the model to achieve statistically significant results. Notably, for larger training samples, the model demonstrates more stable performance, although the marginal gains from additional data become less pronounced. This suggests that while a sufficiently large training sample is crucial for achieving robust generalization, the sample size does not need to be arbitrarily large to achieve significant improvements in model performance, particularly in the context of the testing sample sizes explored in this study.

\begin{table}[h]
\caption{\textbf{Effect of Training Sample Size on Model Performance}
This table reports the Sliced Wasserstein Distance (SWD) between the generated and target distributions using 1 million test events, along with the corresponding permutation-test p-value, for varying training sample sizes. As the training set increases, the SWD tends to decrease and the p-value generally rises, indicating reduced evidence of distributional discrepancy. Once the training sample surpasses roughly 750,000 events, the p-value exceeds 5\%, suggesting that the model’s learned distribution is no longer statistically indistinguishable from the target distribution at conventional significance levels.}
\label{table of train sample}
\tiny
\begin{tabular}{@{}ccc@{}}
\toprule
\textbf{Training Sample (x10$^5$)} & \textbf{SWD test distance (x10$^3$)} & \textbf{SWD test p value} \\
\midrule
18.0   & 0.9551 & 26.6\% \\
15.0   & 0.8893 & 45.3\% \\
12.5   & 1.0309 & 11.1\% \\
10.0   & 0.9356 & 27.9\% \\
\textbf{7.5}   & \textbf{1.0538} & \textbf{7.4\%} \\
5.0    & 1.2293 & 0.7\% \\
2.5    & 1.5182 & 0\% \\
1.0    & 1.5774 & 0\% \\
0.5    & 2.3670 & 0\% \\
\botrule
\end{tabular}
\end{table}

In contrast, suppose that we have only 300,000 training events and wish to generate test events that remain indistinguishable from the target distribution at conventional significance levels.  Table \ref{table of test sample} explores how the p-value changes for varying test sample sizes, identifying the approximate threshold beyond which the model’s outputs become statistically distinguishable. When the test sample size is below 350,000 events, the measured p-value surpasses 5\%, implying that the generated distribution is not distinguishable from the target distribution at conventional significance levels. Once the test set exceeds 350,000 events, however, the p-value falls below 5\%, indicating that the generated and target distributions can be differentiated more confidently.

\begin{table}[ht]
    \caption{\textbf{Test Sample Size vs. Permutation-Test p-Value} This table reports p-values for a model trained on 300,000 events under varying test sample sizes. For smaller test sets (i.e., below 350,000 events), the p-value remains above the 5\% threshold, indicating no statistically significant difference between the generated and target distributions.}
    \centering
    \begin{tabular}{c c}
    \hline
    \textbf{Test Sample (x10$^4$)} & \textbf{P Value (\%)} \\
    \hline
    5  & 60.1\% \\
    10 & 33.3\% \\
    15 & 36.0\% \\
    20 & 7.7\%  \\
    25 & 11.6\% \\
    30 & 5.9\%  \\
    \textbf{35} & \textbf{5.1\%}  \\
    40 & 2.5\%  \\
    45 & 4.7\%  \\
    50 & 3.2\%  \\
    \hline
    \end{tabular}

\label{table of test sample}
\end{table}

The findings presented in this study indicate that achieving statistically indistinguishable generated distributions typically requires a training sample size comparable to the desired testing sample size. This 1:1 ratio suggests that the model effectively captures the distributional characteristics from limited training samples and can reproduce events at the same order of magnitude. An important practical benefit of this approach is significant storage savings. By encoding the Monte Carlo distributional features into a trained model, the storage required is drastically reduced compared to storing large-scale Monte Carlo event samples directly. However, there are limitations inherent in any data-driven method. When the desired generated event sample size significantly exceeds the available training data, there is a risk that the limited training dataset might not sufficiently represent the full distributional space, potentially compromising the reliability of the generated events.

\subsection{Sensitivity Test}

This section examines the performance of three distance metrics-Fréchet Physics Distance (FPD), Kernel Physics Distance (KPD), and Sliced Wasserstein Distance (SWD)-in detecting distributional shifts and covariance alterations within a bivariate Gaussian distribution characterized by zero mean and covariance matrix $\Sigma=\left(\begin{array}{ll}1.00 & 0.25 \\ 0.25 & 1.00\end{array}\right)$ \cite{Kansal_2023}. We evaluate these metrics under a series of distortions, including large and small shifts in the $x$-coordinate and changes in covariance magnitude (multiplying and dividing the variances).

In order to assess the sensitivity of these metrics to subtle biases between the test and target distributions, we introduce very small shifts in the x-coordinate of a bivariate Gaussian. Table \ref{table of shift mu} demonstrate that both FPD and KPD, which are established as baseline methods for evaluation of generative model in High‑Energy Physics (HEP), exhibit high sensitivity to large shifts. These metrics effectively reject the null hypothesis of no mismatch between the true and perturbed distributions, with p-values remaining below $5 \%$ in the presence of substantial distortions. However, when confronted with smaller shifts, particularly when $\mu_x=0.01$ or $\mu_x=0.005$, FPD and KPD demonstrate limitations in their ability to detect distributional mismatches.

\begin{table*}[]
\centering
\caption{Percentage of p-values obtained for various metrics across different noise levels (\(\mu_x\)) with sample sizes of 50,000, 100,000, and 500,000. The underlying distribution is a bivariate Gaussian with zero mean and covariance matrix \(\Sigma=\begin{pmatrix}1.00 & 0.25 \\ 0.25 & 1.00\end{pmatrix}\).}
\begin{tabular}{lcccccc}
\hline
\textbf{Sample Size} & \textbf{Metric} & \textbf{Shift \(\mu_x = 0.1\)} & \textbf{Shift \(\mu_x = 0.05\)} & \textbf{Shift \(\mu_x = 0.01\)} & \textbf{Shift \(\mu_x = 0.005\)} & \textbf{Shift \(\mu_x = 0\)} \\
\hline
 & FPD & 0.0\% & 8.7\% & 96.6\% & 95.4\% & 90.9\% \\
50,000 & KPD & 0.0\% & 2.0\% & 91.1\% & 85.6\% & 91.2\% \\
 & SWD & 0.0\% & 0.0\% & 66.1\% & 61.8\% & 64.5\% \\
\hline
 & FPD & 0.0\% & 0.3\% & 96.0\% & 98.1\% & 92.2\% \\
100,000 & KPD & 0.0\% & 0.0\% & 56.1\% & 45.3\% & 42.7\% \\
 & SWD & 0.0\% & 0.0\% & 19.3\% & 68.7\% & 38.4\% \\
\hline
 & FPD & 0.0\% & 2.0\% & 94.1\% & 99.6\% & 99.9\% \\
500,000 & KPD & 0.0\% & 0.0\% & 79.6\% & 42.3\% & 38.2\% \\
 & SWD & 0.0\% & 0.0\% & 0.0\% & 27.7\% & 25.0\% \\
\hline
\end{tabular}
\label{table of shift mu}
\end{table*}

In contrast, SWD, our novel metric based on the Sliced Wasserstein distance, provides a more stringent test and demonstrates enhanced sensitivity, particularly with smaller shifts and larger sample sizes. For example, when $\mu_x=0.01$, a minimal shift, SWD successfully detects the distributional discrepancy at larger sample sizes, such as 500,000. This is in stark contrast to FPD and KPD, which fail to identify this subtle shift at the same sample sizes. The superior performance of SWD in detecting small shifts is especially pronounced when the sample size is 500,000, where it maintains high detection accuracy. This suggests that SWD is more adept at identifying small distributional differences.

To further examine how small changes in quality and diversity between the test and target distributions impact each metric’s performance, we alter the covariance matrix by multiplying the original values by very minor factors. As shown in Table \ref{table of covariance factor}, FPD and KPD demonstrate relatively consistent performance across different sample sizes when detecting distributional differences under altered covariance conditions. These metrics show no significant sensitivity change with varying sample sizes, though they occasionally fail to identify discrepancies when the covariance is altered, particularly in cases where the covariance factor is multiplied by 1.1 or divided by 0.9. On the other hand, SWD performs well when the covariance changes are more pronounced, other than detecting very subtle differences, such as when the covariance is multiplied by 0.95. Additionally, the performance of SWD improves as the sample size increases, further demonstrating its advantage in detecting distributional mismatches when larger datasets are available.

\begin{table*}[]
\centering
\caption{Percentage of p-values obtained for various metrics across different covariance multiplication factors with sample sizes of 50,000, 100,000, and 500,000. The underlying distribution is a bivariate Gaussian with zero mean and covariance matrix \(\Sigma=\begin{pmatrix}1.00 & 0.25 \\ 0.25 & 1.00\end{pmatrix}\), scaled by the specified factors.}
\begin{tabular}{lcccccc}
\hline
\textbf{Sample Size} & \textbf{Metric} & \textbf{Factor = 0.8} & \textbf{Factor = 0.9} & \textbf{Factor = 1.1} & \textbf{Factor = 1.2} \\
\hline
 & FPD & 20.9\% & 79.4\% & 74.0\% & 3.6\% \\
50,000 & KPD & 71.8\% & 79.2\% & 98.2\% & 93.4\% \\
 & SWD & 0.2\% & 11.4\% & 13.1\% & 0.1\% \\
\hline
 & FPD & 33.1\% & 92.0\% & 75.2\% & 13.7\% \\
100,000 & KPD & 66.6\% & 46.8\% & 61.5\% & 86.0\% \\
 & SWD & 0.0\% & 39.8\% & 0.9\% & 0.0\% \\
\hline
 & FPD & 36.2\% & 89.0\% & 73.5\% & 22.3\% \\
500,000 & KPD & 82.4\% & 65.0\% & 50.1\% & 26.4\% \\
 & SWD & 0.0\% & 0.0\% & 0.0\% & 0.0\% \\
\hline
\end{tabular}
\label{table of covariance factor}
\end{table*}

In conclusion, while FPD and KPD provide reliable detection of larger shifts and major covariance changes, SWD emerges as a more sensitive and stringent metric, particularly when dealing with small shifts and covariance alterations. The performance of all metrics improves as the sample size increases, with SWD showing consistent superiority in detecting subtle differences in data distributions. These findings suggest that SWD is particularly beneficial in situations requiring high sensitivity to small distributional changes, making it a more precise method for evaluating generative models. In the context of High-Energy Physics (HEP), where accurate modeling and evaluation of generative models are crucial, SWD offers a more stringent approach for detecting discrepancies in data distributions, especially when high precision is required. Therefore, for tasks that demand high precision in the evaluation of generative models, we recommend using the SWD method, as it consistently provides superior detection capabilities even with small deviations in data.

\section{Conclusion}
In conclusion, we find that the Sliced Wasserstein Distance can serve as an efficient and universal loss function for measuring the distribution consistency in high-dimensional data. Our generative model makes it difficult to manually distinguish distribution consistency based on histogram differences in specific dimensions or arbitrary random projections. This result indicates that our model has high accuracy and reliability in simulating the distribution characteristics of high-dimensional data. Our analysis demonstrates that beyond a training sample size of roughly the same order of magnitude in event counts, the generative model consistently achieves statistically indistinguishable results from the target distribution. We used the sliced Wasserstein distance and the permutation test to calculate the confidence level, providing a rigorous and more sensitive method for measuring the consistency of the distribution. Additionally, our comparative sensitivity studies confirm that SWD provides a superior capability to detect subtle shifts and covariance alterations compared to traditional metrics such as FPD and KPD, highlighting its utility for precision-critical assessments in high-energy physics generative modeling. The working principle and advantages of this method are that it can provide a quantitative assessment of distribution consistency. The data that support the findings of this article are openly available\cite{pan_2023_10066889}. Our analysis code and results are publicly available on GitHub at \url{https://github.com/caihao/SWD-EvtGen} to support open science and reproducibility. For any questions or issues, please raise an issue in the repository or contact us directly.

\bibliography{apssamp}

\end{document}